\documentstyle[aps,prl,epsfig,psfig,twoside,graphicx,preprint]{revtex}

\tightenlines
\begin{document}

\title{Anomalous self-similarity in a turbulent rapidly rotating fluid}

\author{Charles N. Baroud$^1$, Brendan B. Plapp$^1$, Zhen-Su She$^{2,3}$,
and Harry L. Swinney$^1$} 

\address{$^1$Center for Nonlinear Dynamics and
Department of Physics, The University of Texas at Austin, Austin, TX
78712\\
$^2$State Key Laboratory for Turbulence Research, Department of Mechanics
and Engineering Science, Peking University, Beijing 100871, China\\
$^3$Department of Mathematics, UCLA, Los Angeles, CA 90095}

\date{\today} 
\maketitle

\begin{abstract} 
Our velocity measurements on the quasi-two-dimensional turbulent flow in
a rapidly rotating annulus yield an inverse energy cascade with $E(k)
\sim k^{-2}$ rather than the expected $E(k) \sim k^{-5/3}$. The
probability distribution functions for longitudinal velocity
differences, $\delta v(\ell) = v(x+\ell)-v(x)$, are self-similar (scale
independent) but strongly non-Gaussian, which suggests that the coherent
vortices play a significant role.  The structure functions,
\protect{$\langle[\delta v(\ell)]^p\rangle\sim \ell^{\zeta_p}$}, exhibit
anomalous scaling: $\zeta_p=\frac{p}{2}$ rather than the expected
$\zeta_p=\frac{p}{3}$. 
\end{abstract}

\draft

In large scale flows in the earth's atmosphere and oceans or in gaseous
planets, the Coriolis force dominates other forces and to lowest order
is balanced by pressure gradients (geostrophic balance).  The
dimensionless number that characterizes this regime is the Rossby
number, the ratio of magnitudes of the inertial term in the
Navier-Stokes equation ($\vec{u} \cdot \nabla \vec{u}$, where $\vec{u}$
is the velocity) to the Coriolis term ($2 \vec{\Omega} \times \vec{u}$,
where $\Omega$ is the rotation rate). For planetary flows on large
scales the Rossby number is typically in the range 0.05-2, and
turbulence in such flows is quite different from three-dimensional (3D)
turbulence in an inertial reference frame.  Despite the importance of
planetary flows, geostrophic turbulence has scarcely been examined in
laboratory experiments.  Our experiments in a rotating annulus are the
first to determine the scaling properties of turbulence in a low Rossby
number flow.
 
One of the most significant effects of rapid rotation on a fluid is the
two-dimensionalization of the flow~\cite{frisch}. Recent numerical
simulations of a rotating turbulent flow recovered
quasi-2-dimensionality even in the case of 3D
forcing~\cite{smith99}. Turbulence in a 2D rapidly rotating flow has
some similarities to other 2D turbulent flows, but the rapid rotation
leads to the formation of large vortices and jets, hence a flow that is
more inhomogeneous and anisotropic~\cite{frisch}.

The statistics of velocity differences for 2D turbulence are predicted
to be self-similar~\cite{Siggia81}, i.e., the probability distribution
function (PDF) for the difference between velocities measured at two
points, $\delta v(\ell) = v(x+\ell)-v(x)$ (where $v$ is along the line
connecting the two points), should have a functional form independent of
the separation $\ell$. Self-similarity in a turbulent flow can also be
determined from the scaling of structure functions,
$S_{p}(\ell)\equiv<[\delta v(\ell)]^p> \sim \ell^{\zeta_p}$: for any
self-similar flow, $\zeta_p$ will vary linearly with
$p$~\cite{frisch}. In particular, Kolmogorov's 1941 theory for
homogeneous isotropic turbulence predicts the existence of an inertial
range with $\zeta_p = {\frac{p}{3}}$.

Experiments and numerical simulations have shown that 3D turbulence is
$\em{not}$ self-similar~\cite{frisch}. The observed deviations from a
linear dependence of $\zeta_p$ on $p$ are attributed to the stretching
and folding of vortices. As a vortex stretches, it shrinks in lateral
extent until it collapses onto a singular line. Consequently, the energy
dissipation rate varies with length scale. Vortex stretching is not
allowed in 2D turbulence, and recent numerical simulations of 2D
turbulence yielded self-similar behavior~\cite{boffetta00} (see also
~\cite{Siggia81,smith93}). A recent experiment yielded self-similar
behavior with a Gaussian PDF~\cite{paret98} while two other experiments
showed scale-dependent statistics~\cite{vorobieff99,daniel00}, as will
be discussed later.

Our experiments on turbulent flow in a rapidly rotating annulus yield a
self-similar PDF for the velocity differences, but the PDF is not
Gaussian and the structure function exponent values $\zeta_{p}$ differ
from the expected $p/3$. We now describe our experiments and then
present the results.

Our apparatus consists of an annular tank filled with water and covered
with a solid lid; the inner diameter of the tank is 21.6 cm and the
outer diameter is 86.4 cm ~\cite{solomon92}. The depth of the tank
increases from 17.1~cm at the inner radius to 20.3~cm at the outer
radius~\cite{beta}. Flow in the annulus is produced by continuously
pumping water in a closed circuit through two concentric rings of 120
holes each in the bottom of the tank; the source ring is at a radius of
18.9 cm and the sink ring is at 35.1 cm.  The radially outward flux from
the pumping couples with the Coriolis force to generate a
counter-rotating jet, which is wide and turbulent over a wide range of
parameters. The rapid rotation leads to a flow that is essentially 2D
except in the thin Ekman boundary layers at the top and bottom
surfaces~\cite{solomon92}.

In the present experiments the tank rotates at 11.0~rad/s, sufficiently
fast to produce an essentially 2D flow, and the flux is 150~cm$^3$/s,
sufficiently large to produce an inverse energy cascade. The azimuthal
velocity is measured using two hot film probes which are inserted
through the top lid and extend 1~cm into the water on opposite sides of
the tank, midway between the inner and outer walls. Each probe was
sampled at 150~Hz for periods of two hours, giving $10^6$ data points
per probe for each run, and the measurements were repeated four times,
yielding a total of $8\times10^6$ data points. Using the maximum
velocity ($U_{max}=22$~cm/s) as the velocity scale and the distance
between the forcing rings ($L=16.2$~cm) as the integral length scale, we
calculate the Reynolds number to be $35,000$ and the Rossby number
($U_{max}/2\Omega L$) to be $0.06$.

Instantaneous 2D velocity fields were obtained using a Particle Image
Velocimetry (PIV) system with a horizontal light sheet at mid-fluid
depth and a rotating camera above the tank. For each flow condition, 50
instantaneous velocity fields were obtained, equivalent to approximately
$2\times10^5$ velocity values at the radius of the hot film
probes. Though this sample size is inadequate for high order statistics,
the spatial information provided by the PIV measurements complements the
long velocity time series obtained with the hot film probes. We also
made PIV measurements at the same rotation rate at a higher pumping rate
of 550~cm$^3$/s ($Re=100,000$).

\begin{figure}[b]
\includegraphics[width=\linewidth]{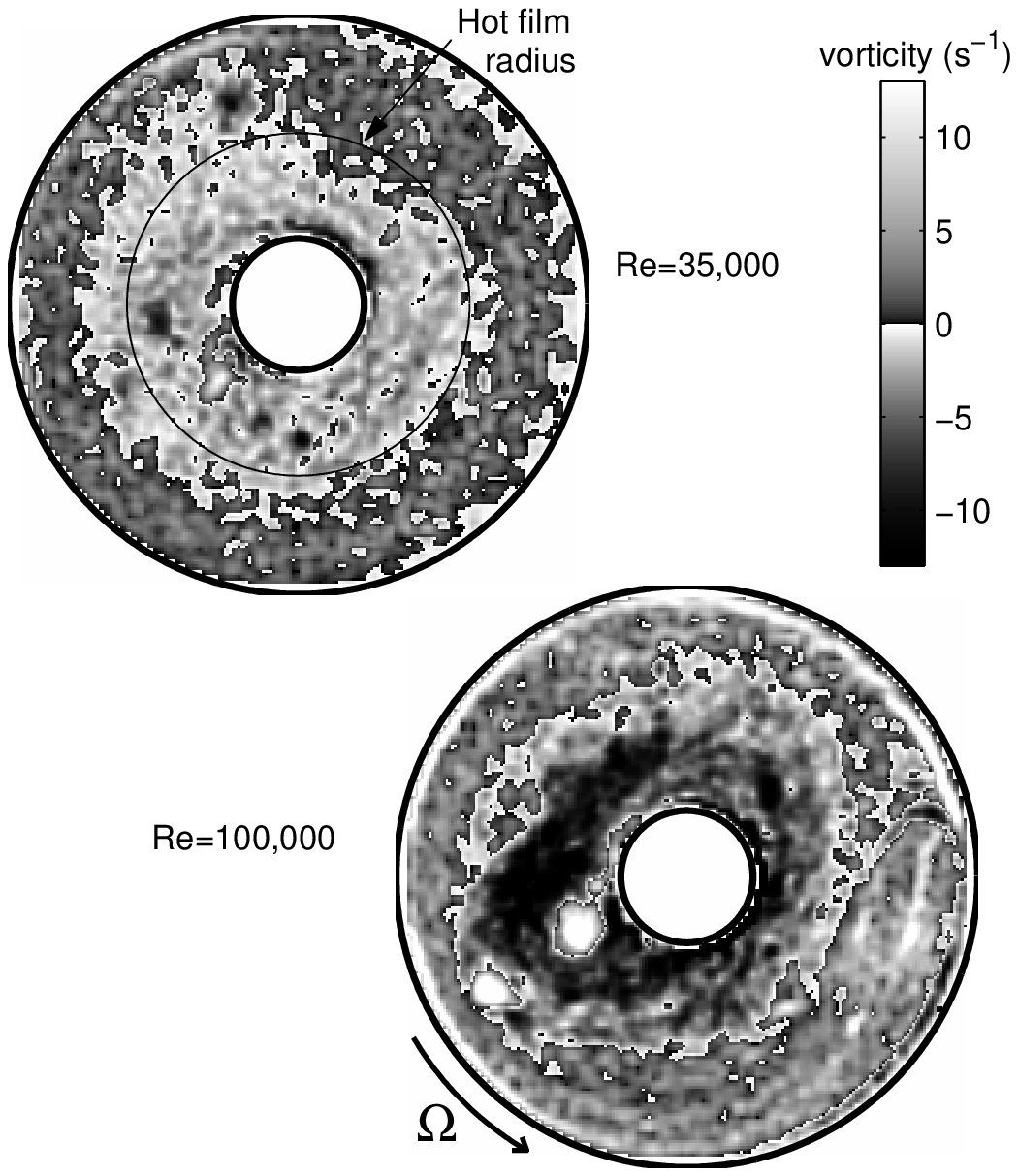}
\caption{Vorticity field at Reynolds number 35,000 ($Q=150$ cm$^3$/s)
and 100,000 ($Q=550$ cm$^3$/s) for rotation rate $11.0$ rad/s.  Vortices
with light (dark) center are cyclonic (anticyclonic).  The vortices are
advected by the clockwise jet. }
\label{fig:field}
\end{figure}

Two vorticity profiles obtained with the PIV system are shown in
Fig.~\ref{fig:field}.  When the pump is first turned on, the flow
consists of rings of cyclonic and anti-cyclonic vortices that form above
the outlet and inlet rings, respectively. Vortices of like sign
immediately begin merging and growing in size, and a counter-rotating
jet forms between the two rings. At long times the inverse energy
cascade leads to large vortices which are larger at higher pumping
rates, as Fig.~\ref{fig:field} illustrates.

\begin{figure}[htp!]
\includegraphics[width=\linewidth]{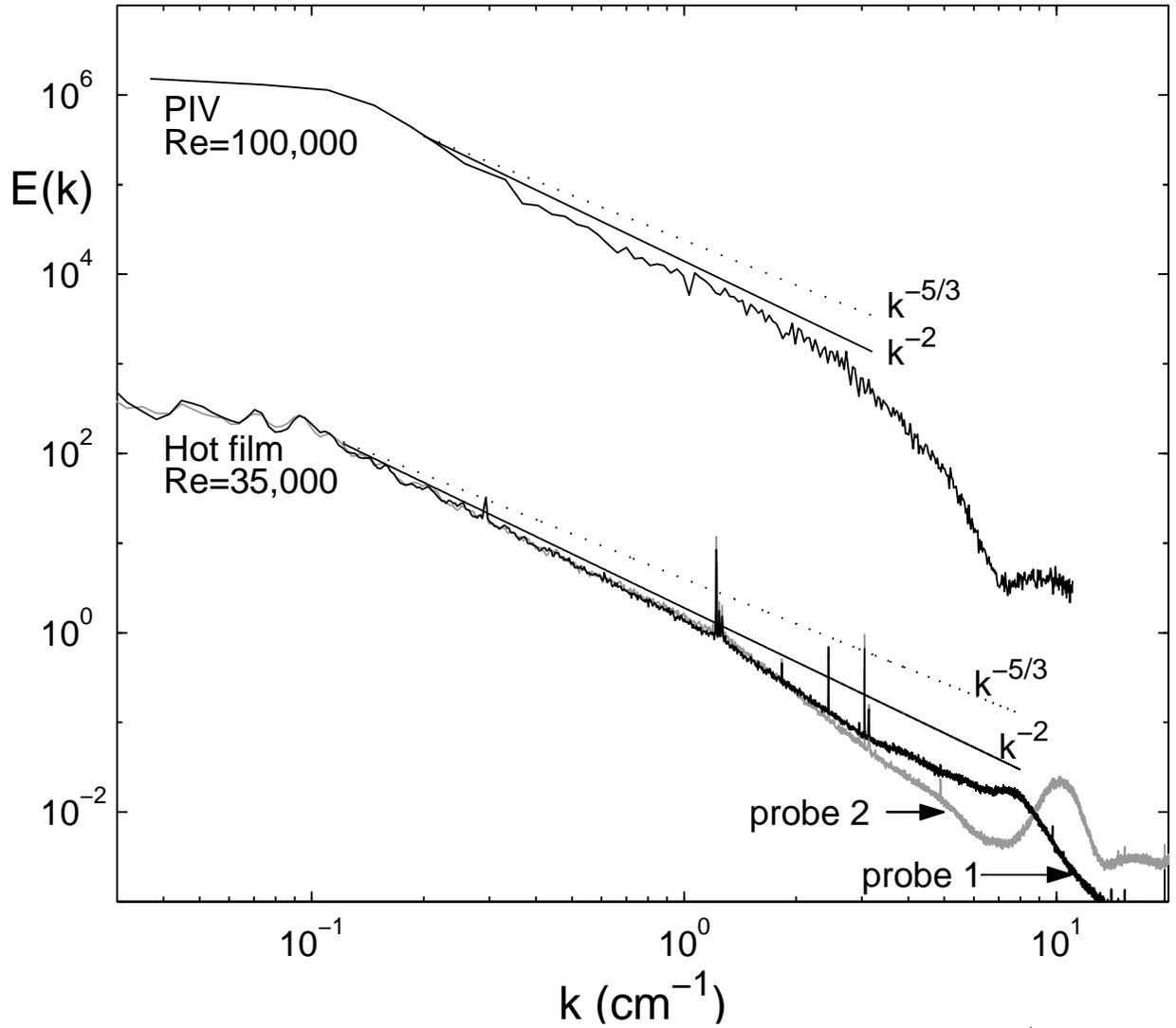}
\caption{Energy spectra (arb. units) with a dotted line showing the
Kraichnan $k^{-5/3}$ inverse cascade and a solid line showing $k^{-2}$
behavior. Fits of all of the hot film spectra in the range
$0.1<k<1.22$~cm$^{-1}$ give a slope of $-2.04\pm 0.06$. The sharp
spectral peaks correspond to harmonics of the tank's rotation rate, not
to dynamics of the flow.}
\label{fig:spectra}
\end{figure}

We compute energy power spectra from the velocity time series data
assuming Taylor's frozen turbulence hypothesis, which is applicable
because the turbulent intensity (ratio of the rms velocity fluctuation
to the mean velocity) is less than 10\%. The spectra contain a region
with $E(k)\sim k^{-2}$ (Fig.~\ref{fig:spectra}), in contrast with
Kraichnan's prediction of $E(k)\sim k^{-5/3}$ for the inverse
cascade~\cite{kraichnan67}. Energy spectra obtained from PIV
measurements of the azimuthal velocity data at $Re=100,000$ show a
scaling consistent with that obtained from the time series data, as
shown in the upper curve of Fig.~\ref{fig:spectra}. A prediction of
$E(k)\sim k^{-2}$ has been made for a turbulent flow made up of Lundgren
spiral vortices that were not allowed to stretch in the third
dimension~\cite{gilbert88}. Such scaling has also been predicted for
rotating turbulent flows as they become 2D~\cite{zhou95}.

At high wavenumbers, our power spectra are consistent with those
previously found for the forward (enstrophy) cascade, $E(k)\sim k^{-n}$,
where $3\le n\le4$. However, our spectral range is too small to deduce a
value for $n$.

\begin{figure}
\includegraphics[width=\linewidth]{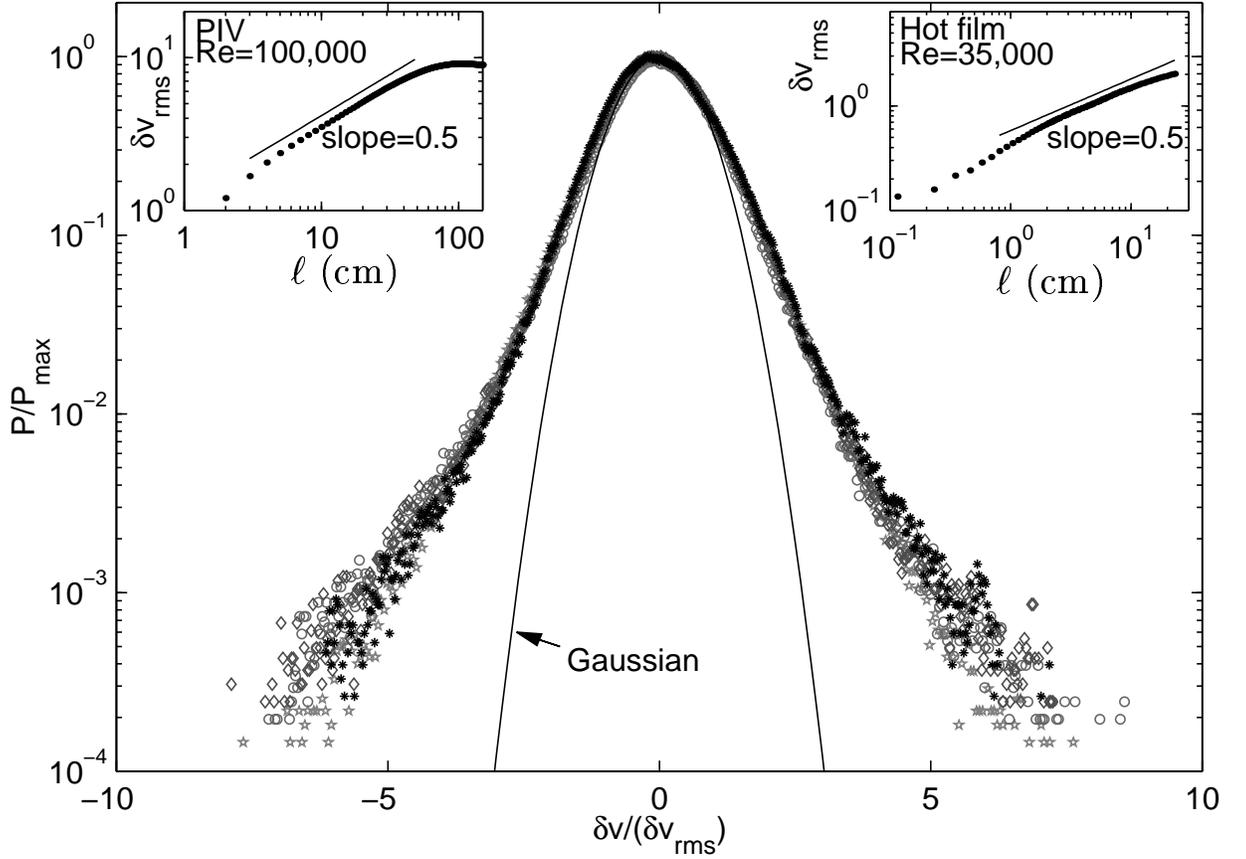}
\caption{Normalized probability distribution function for the velocity
differences, demonstrating self-similar behavior: data for different
separations ($\ell=0.6,4.6,9.2,17.3$~cm) collapse onto a single curve.
The velocity differences are normalized by their standard deviation
$(\delta v)_{rms}$ and the probability by its maximum value,
$P_{max}$. The standard deviation of the velocity differences (see
insets) scales as $\ell^\alpha$ where $\alpha=0.50\pm 0.06$ for
$Re=35,000$ and $\alpha=0.54\pm 0.06$ for $Re=100,000$.}
\label{fig:pdfs}
\end{figure}

By plotting the PDF for the velocity differences $\delta v(\ell)$
(Fig.~\ref{fig:pdfs}), we obtain the first indication of
self-similarity: data for different separations $\ell$ collapse onto a
single curve when normalized. This curve is far from Gaussian, and the
enhanced probability in the tails is likely due to the strong velocity
differences that arise as coherent vortices pass the probes, as noted by
She et al.~\cite{she91}.

Another indication of anomalous behavior is the scaling of the standard
deviation of the velocity differences, $\delta v_{rms}$. Our data
suggest $\delta v_{rms}(\ell)\sim \ell^{1/2}$ (see insets of
Fig.~\ref{fig:pdfs}) rather than the expected $\delta v_{rms}(\ell)\sim
\ell^{1/3}$; the observed scaling of $\delta v_{rms}$ is consistent
with the scaling of the energy spectrum in the inverse cascade range
(Fig.~\ref{fig:spectra}).

\begin{figure}
\includegraphics[width=\linewidth]{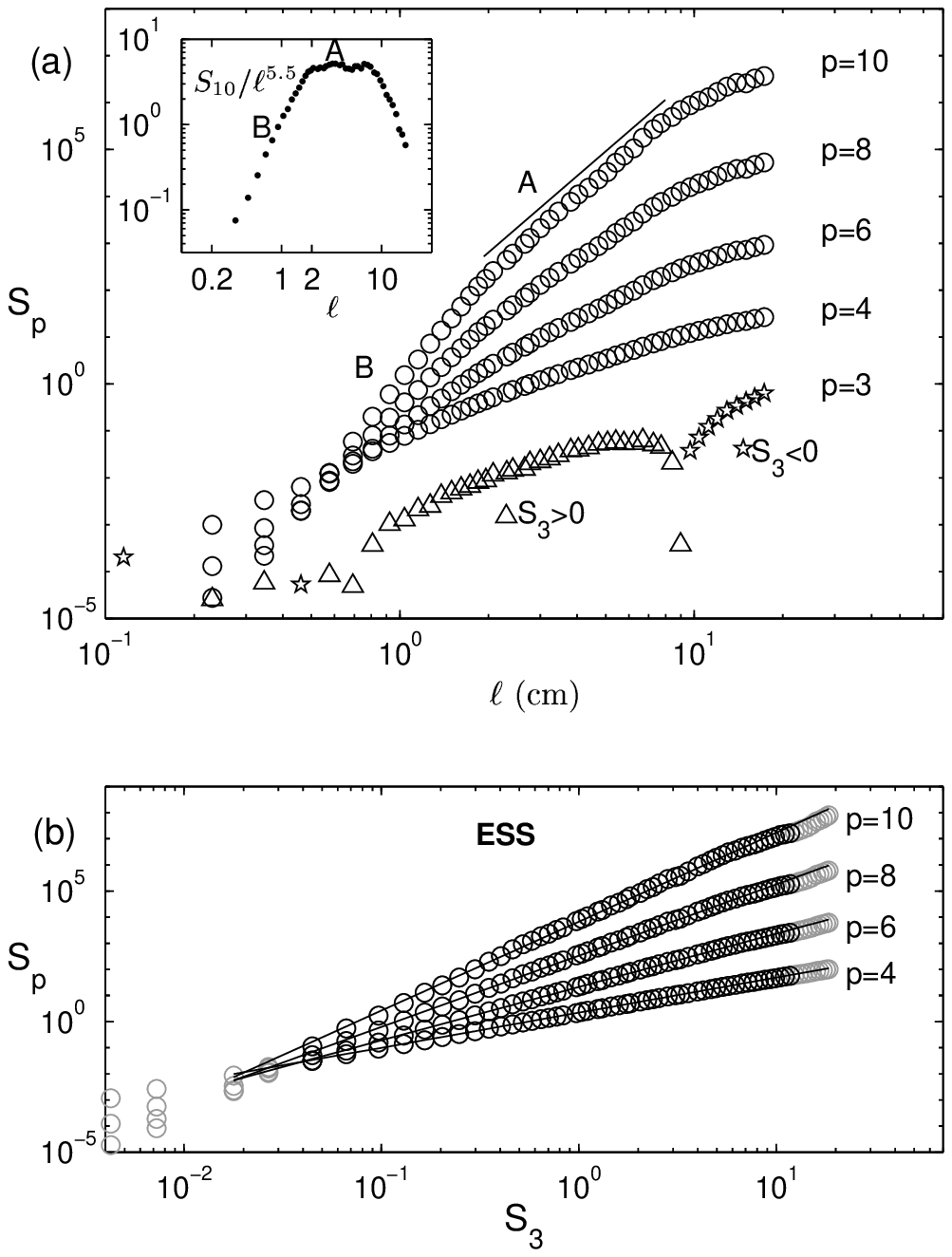}
\caption{Even order structure functions (a) as a function of $\ell$ and
(b) as a function of $S_3$ (an Extended Self Similarity plot), for the
hot film data at $Re=35,000$. The graph of $S_{10}/\ell^{5.5}$ in the
inset in (a) emphasizes the sharpness of the bend in $S_{10}$ at $\ell
\simeq 2$ cm.  In (b) the dark symbols indicate the region from which
the values of $\zeta_p$ were extracted.}
\label{fig:sf-l}
\end{figure}

Now we examine the structure function scaling, plotting $S_p$ as a
function of $\ell$, as shown in Fig.~\ref{fig:sf-l}(a). There is a
scaling region (labeled A) from about 2~cm to 8~cm, and some indication
of another scaling region (labeled B) for $\ell<2$ cm; the transition
between the two regions can be seen more clearly in the plot of $S_{10}
/\ell^{5.5}$ vs. $\ell$ in the inset. The 2 cm lower limit of scaling of
the structure functions is below the 5 cm ($k=1.2$~cm$^{-1}$ in
Fig.~\ref{fig:spectra}) length where the energy spectrum behavior
changes; this difference might be due to systematic error in calibration
of the probes (note in Fig. 2 the difference in two probes at high
wavenumbers), or perhaps because the scaling regions in real space and
in Fourier space do not exactly correspond, as discussed in
Ref.~\cite{frisch} (p.~62).

One test for the existence of an inverse energy cascade in 2D turbulence
is the sign of $S_3(\ell)$: Kolmogorov's four-fifths law can be written
for anisotropic flows as $\varepsilon = -\frac{1}{4}\nabla_\ell \cdot
\langle|\delta {\bf v}({\bf \ell})|^2\delta {\bf v}({\bf \ell})
\rangle$, where ${\bf v}$ is the full 3D velocity vector and
$\varepsilon$ is the mean rate of energy transfer (Ref. ~\cite{frisch},
p.~88). For the 3D forward cascade $\varepsilon$ is positive; a negative
value of $\varepsilon$ corresponds to an inverse cascade. If the
anisotropy is not too strong, the longitudinal $S_3$ dominates the
transverse $S_3$, and one can obtain the sign of $\varepsilon$ from
measurements of the longitudinal structure function alone. While our
flows are anisotropic, we find that PDFs for velocity differences from
PIV measurements in the radial and azimuthal direction are in agreement,
indicating that the anisotropy does not strongly affect the velocity
differences. The hot film data yield $S_3>0$ for $\ell<10$~cm, which
suggests that the inverse energy cascade stops at that point. This
length corresponds visually to the scale of the largest vortices; thus
region A corresponds to the inverse cascade. Scales below about 1~cm
(the distance between the holes in the inner ring) presumably correspond
to the forward enstrophy cascade.

The now standard way of extracting structure function exponents from
data with limited inertial range is the technique introduced by Benzi et
al.~\cite{benzietal} called Extended Self-Similarity (ESS), where the
$\zeta_p$ values are obtained from the slope of log-log plots of $S_p$
vs. $S_3$. The ESS plots of our data, Fig.~\ref{fig:sf-l}(b), exhibit
power law scaling through both ranges A and B, for lengths
$0.5<\ell<15$~cm. Since $S_3$ should contain the same trends as the
other structure functions, it is not surprising that the switching from
region A to B is lost by this method.

\begin{figure}
\includegraphics[width=\linewidth]{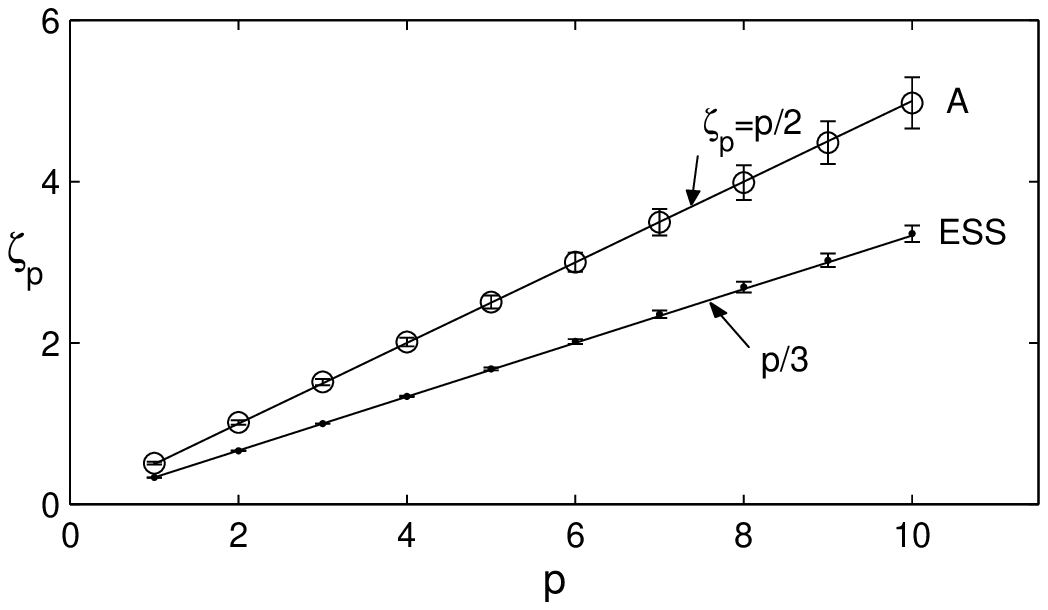}
\caption{Scaling exponents $\zeta_p$ as a function of $p$ for region A
of Fig.~\ref{fig:sf-l}(a), and from the Extended Self Similarity
analysis, Fig.~\ref{fig:sf-l}(b).  Region A, the inverse energy cascade
region, yields $\zeta_p=(0.50\pm 0.03)p$, while the ESS scaling shows
the expected $\zeta_p=(0.33\pm 0.01)p$ scaling over the forward and
inverse cascade ranges. The error bars show the standard deviation of
the eight separate data sets of $10^6$ points.}
\label{fig:two-slopes}
\end{figure}

The exponent values $\zeta_p$ deduced from the plot of $S_p(\ell)$
[region A in Fig.~\ref{fig:sf-l}(a)] and from the ESS plot
[Fig.~\ref{fig:sf-l}(b)] are compared in
Fig.~\ref{fig:two-slopes}. Region A of Fig.~\ref{fig:sf-l}(a) yields
$\zeta_{p,A}=p/2$, in contrast with theory, although $\zeta_{p,ESS} =
p/3$, as it must for a self-similar flow.

Two recent experiments on turbulence in a quasi-2D fluid, a soap film,
found departures from linear scaling of
$\zeta_p$~vs.~$p$~\cite{vorobieff99,daniel00}, indicating that these
flows were not self-similar. Another experiment on a quasi-2D flow, a
magnetically driven shallow layer of electrolyte, yielded an inverse
energy cascade with self-similar behavior as indicated by both the PDFs
for $\delta v$ and by the scaling of the structure function computed
using ESS, which scaled as $\zeta_p=p/3$~\cite{paret98}; however, the
PDFs for $\delta v$ were close to Gaussian, in contrast to those yielded
by our flow (Fig.~\ref{fig:pdfs}). A recent numerical simulation of 2D
turbulence also yielded self-similar PDFs for $\delta v$, where the
departure from Gaussianity, though small, contained significant
information about the flow~\cite{boffetta00}.

The $p/2$ scaling obtained for $\zeta_p$ in our flow may be a
consequence of an inverse cascade driven by a radial velocity
shear. This process would yield an energy flux $\varepsilon\sim \delta
v_s (\delta v_{\parallel})^2/\ell$, where $v_s$ would be a radial
velocity due to the interaction of the vortices with the mean shear,
$v_{\parallel}$ the azimuthal velocity and $\ell$ the azimuthal
separation. For a scale-invariant $\varepsilon$, the resulting $p^{th}$
order velocity structure function would have the scaling exponent $p/2$.

In summary, we find that turbulence in our rapidly rotating flow is
self-similar, as indicated by the collapse of the PDFs onto a single
curve and by the linear scaling of the structure function
exponents. This self similarity extends over both the inverse and
forward cascades. In the inverse cascade range, the flow exhibits
anomalous scaling of the energy, $E(k)\sim k^{-2}$, and the structure
functions, $S_p\sim\ell^{p/2}$, but this anomalous scaling is missed in
an Extended Self Similarity analysis. We conclude that the rotation and
strong shear make this flow different from 2D turbulent flows in
non-rotating systems. Our results suggest that the transport and mixing
in low Rossby number flows such as the atmosphere and oceans may be
different from these processes in isotropic homogeneous 2D turbulent
flows.

This research was supported by a grant from the Office of Naval
Research. Z-S. S. acknowledges partial support from the Minister of
Education in China and the Natural Science Foundation of China.

\begin{thebibliography}{99}
\bibitem{beta} 
The sloping bottom, a ``beta plane'', approximates to first order the
variation of the Coriolis force with latitude. The role of the beta
plane, which breaks the symmetry between co-rotating and
counter-rotating jets, will be the subject of a future study.
\end{thebibliography}
\end{document}